# Solar Enablement Initiative in Australia: Report on Efficiently Identifying Critical Cases for Evaluating the Voltage Impact of Large PV Investment

Mehdi Shafiei, Aaron Liu, Gerard Ledwich, Geoffery Walker
School of Electrical Engineering and Computer Science
Queensland University of Technology (QUT)
Brisbane, Australia
m.shafiei@qut.edu.au, lei.liu@connect.qut.edu.au,
g.ledwich@qut.edu.au, geoffrey.walker@qut.edu.au

Gian-Marco Morosini, Jack Terry
School of Information Technology and Electrical Engineering
The University of Queensland (UQ)
Brisbane, Australia
g.morosini@uq.edu.au, jack.terry@uq.edu.au

*Abstract*— **The increasing quantity of PV generation connected to distribution networks is creating challenges in maintaining and controlling voltages in those distribution networks. Determining the maximum hosting capacity for new PV installations based on the historical data is an essential task for distribution networks. Analyzing all historical data in large distribution networks is impractical. Therefore, this paper focuses on how to time efficiently identify the critical cases for evaluating the voltage impacts of the new large PV applications in medium voltage (MV) distribution networks. A systematic approach is proposed to cluster medium voltage nodes based on electrical adjacency and time blocks. MV nodes are clustered along with the voltage magnitudes and time blocks. Critical cases of each cluster can be used for further power flow study. This method is scalable and can time efficiently identify cases for evaluating PV investment on medium voltage networks.**

*Index Terms*—medium voltage distribution network, photovoltaic, voltage rise, K-means clustering, Gaussian distribution

## I. Introduction

The distribution network PV penetration is high in Australia, with 32% of dwellings in the state of Queensland with PV in September 2018 [1] and some communities have greater than 50% of houses with rooftop PV [2]. With government incentives, increasing electricity prices and falling PV prices, more customers want to increase control over their electricity supply by installing PV [3]. However, a high penetration percentage of rooftop PVs can result in overvoltage issues [4]. Typically, distribution network service providers (DNSPs) have a manual process to assess the voltage impact of PV investment applications. The existing process runs power flow analyses on the highest voltage magnitude scenario of a particular node where the new large PV system or aggregated of customers' rooftop PVs is going to be connected to find the limit of PV penetration. There are two major drawbacks of this approach. First, the highest voltage case of the medium voltage (MV) feeder network may not happen on the same node where the new PV is going to be connected. Secondly, there is no time-varying feature in the assessment, while customer loads and PV generations vary with time. In order to address these drawbacks, the solar enablement initiative (SEI) project was initialized in Australia to support DNSPs with a better understanding of the operational conditions of their networks.

Several articles [5-7] have addressed the impact of high PV penetration in distribution networks. To find the maximum hosting capacity of new lumped rooftop PVs in the distribution networks, [8] proposed a new probabilistic method considering the distribution of customer loads and PV generations. However, with the current technologies in the distribution networks, the data of the customer loads are mostly unavailable. In other words, the metered data are net demand data, which provides data of customer load minus local generation. Additionally, there are often insufficient measurements in the distribution networks because there are not sufficient meters with communication capability in the distribution networks.

Finding the maximum hosting capacity of the lumped rooftop PVs requires a comprehensive study on a predefined period of time. Determining this period is of vital importance. For instance, in [9], one-week voltage measurements at 10-min intervals are considered, while one week is very short for capturing the PV outputs and load variations. PV generation has a strong seasonal variation which is strongly correlated with the variation of solar radiation across a year. Hence, a complete season of data may be comprehensive enough for finding the maximum hosting capacity of the lumped rooftop PVs in distribution feeders. In large distribution networks, it is impractical to study all historical time series data. Therefore, one of the critical parts of the SEI project is to identify the critical cases for assessing the limit of PV penetration. Clustering techniques can be efficient in identifying typical samples of a population. In [10, 11], K-means as one of the most popular clustering algorithms is employed to find the critical cases for quantifying the effect of the renewable generation in distribution networks. However, in large distribution networks with extensive data, pre-analysis methods are required to reduce the amount of data to have a small-size data for K-means clustering.

This work aims to develop an accurate algorithm for identifying critical cases to determine the maximum hosting of the lumped rooftop PVs in distribution networks. In this method, a bad data detection band is proposed to remove false data from the available historical data. Nodal grouping is another important step of this algorithm, where the data from neighbouring nodes with similar load type are merged into the same group. Our empirical research shows that a Gaussian distribution can be fitted to the nodal voltage values, which can identify the area of critical cases for PV analysis study. In the last step, the elbow method [12] is employed to find the optimal number of clusters for the two-dimension K-means clustering method.

This paper reports how to cluster MV nodes and identify critical cases time-efficiently for further evaluation of large PV applications. The developed method and results are in the next sections, followed by conclusion and references.

## II. METHOD

In this part, we present a new data processing approach to find the critical cases for determining the maximum allowable of the lumped PV penetration in MV distribution feeders. The developed method consists of several steps to find the minimum number of critical cases, which are comprehensive enough for PV analysis as follow:

### A. Bad data detection

In distribution networks, a combination of the measured and the estimated data are available, while both datasets may contain bad data. For the first step of the developed algorithm, bad data are identified and removed. Our investigations show a Gaussian distribution with an acceptable error can be fitted on MV nodes voltage values (in the study period) with a mean ($\mu_v$) and a standard deviation ($\sigma_v$). For a normally distributed random variable, 99.7% of the nodes' voltage values lie within the band of $\mu_v \pm 3\sigma_v$. Usually, data out of this range can be considered as bad data. However, to avoid removing important information from these datasets, $\mu_v \pm 7\sigma_v$ is considered as a reliable marginal band for bad data [13].

### B. Nodal Grouping-Voltage Differences

In MV distribution networks, the voltage differences between neighbouring nodes are very small. Hence, neighbouring nodes without substantial voltage differences are considered in the same groups, and the critical cases in each group are considered for PV analysis. Fig. 1 represents a graphical algorithm for nodal grouping in MV distribution networks. In this algorithm, $i$ and $j$ represent nodes, $\Delta V_{ij}$ is the voltage difference between nodes $i$ and $j$, and $V_{base}$ is the nominal operating voltage of MV network. The limit for the $\Delta V_{ij}$ represents the range of nodal voltage grouping. In this article, the measurement error is considered as the limit of $\Delta V_{ij}$. This measurement error is between 0.2% to 0.5% of $V_{base}$ [14].

### C. Nodal Grouping-Customer loads Correlation Analysis

Nodal grouping based on the voltage differences between neighbouring nodes is not comprehensive enough in the distribution networks with different types of customer loads. Hence, the cross-correlation between customer loads in each group needs to be considered as a part of the nodal grouping algorithm. A customer load with a low cross-correlation with other group members needs to be considered in a separate group. The cross-correlation coefficient is usually defined mathematically as in (1) [15]:

$$Correlation_{Spatial} = {cov(x,y)}/{\sigma_x * \sigma_y} \qquad (1)$$

where $cov$ and $\sigma$ denote covariance and standard deviation, respectively. For cross-correlation, $x$ and $y$ represent two sets of historical data from two different nodes.

### D. Gaussian Distribution Analysis

Nodal grouping puts neighbouring nodes with a similar load type in the same group. Our empirical investigations show that a Gaussian distribution with a negligible error can be fitted on the nodal voltage values in the same group. Fig. 2 shows a typical Gaussian distribution with mean ($\mu$) and standard deviation ($\sigma$). Voltage values greater than $\mu + 2\sigma$ are critical cases for PV analysis study.

To determine the maximum limit for the new lumped PV generation in MV distribution networks, the critical cases in the upper tail of the fitted Gaussian distribution are considered.

### E. Two Dimensional K-Means Clustering Based on the Voltage Magnitudes and Time

In this part, two dimensional K-means clustering method is employed to represent several sets of data at different voltage levels and time windows of a day. Let $X = \{V, T\}$ be the set of two dimensional points to be clustered into $K$ clusters, with centroids $C = \{c_i, i = 1, \dots, k\}$. Implementation of K-means clustering is explained step-by-step as follows [16]:

- **Step 1:** Randomly choose the $K$ number of cluster centroids.
- **Step 2:** Calculate Euclidean distance between data and their centroid as:

$$\arg\{\min_{c_i \in C} dist(c_i, X)^2\} \qquad (2)$$

where $dist(.)$ is Euclidean distance.
- **Step 3:** Find the new centroids by calculating the average of all assigned points to each cluster as:

$$C_i = avg(X_i) \qquad (3)$$

where $X_i$ is the set of the data assigned to cluster $i$.
- **Step 4:** Repeat **Step 2** and **Step 3** until zeros changes in all clusters.

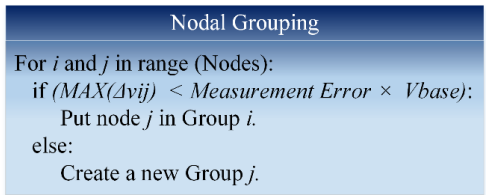

Fig. 1. Nodal grouping.

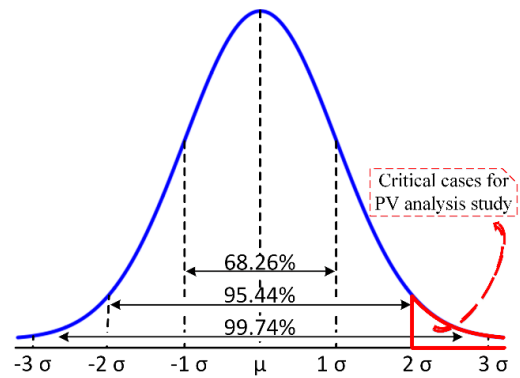

Fig. 2. A typical Gaussian distribution.

K-means clustering by itself is somewhat naive because it cannot find the optimal number of clusters. For this study with different sets of data, the optimal number of clusters is required. Hence, an Elbow method is employed in this section to find the optimal number of clusters [12]. In this method, for $\{K_i, i = 1, 2, ..., k\}$, calculate the sum of squared error (SSE). By increasing the number of the clusters, the SSE tends to decrease towards zero (each data point has its own cluster). The aim of this method is to choose an optimal value for $K$, while the SSE is low enough. Fig. 3 plots a line chart of the SSE for $K_i$. If the line chart is look like an arm, the "Elbow" on the arm is the optimal value for $K$ (as shown in Fig. 3). It is expected that the elbow point may not be clear when the error line chart is smooth. In this situation, the SSE with one cluster (SSE1) is considered as the maximum error, and $K_i$ that gives the SSE less than 30% of SSE1 is chosen as the optimal number (as shown in Fig. 3).

## III. SIMULATION RESULTS

In this section, the developed algorithm is employed to find the critical cases of a real Australian MV distribution network. It is worth noting that the period of the study depends on the seasonal variations in customer loads behaviour. For this research, seasonally based clustering is considered. The measured and estimated data in the summer season are considered in this section. Ten minutes interval measurements, estimated nodal voltages, and injected currents are available for this network. The results of the pre-described data processing steps in Section II are represented below:

### A. Bad data detection

As noted before, bad data in both measured and estimated data decrease the accuracy of the proposed method. Hence, for the first step, we remove bad data to have a more reliable set of data for the next data processing steps. Fig. 4 represents the quantile-quantile (QQ) of the set of data. As shown in this figure, the data out of the $\mu_v \pm 7\sigma_v$ band are labeled as bad data and removed.

### B. Nodal Grouping

Fig. 5 represents a MV distribution network with 49 nodes and 16 MV/LV transformers with customer loads. For the first step of nodal grouping, based on the available data, the voltage differences matrix is calculated and visualized in Fig. 6 (a). 0.2% of the network voltage base (11 kV) is considered as the threshold for grouping.

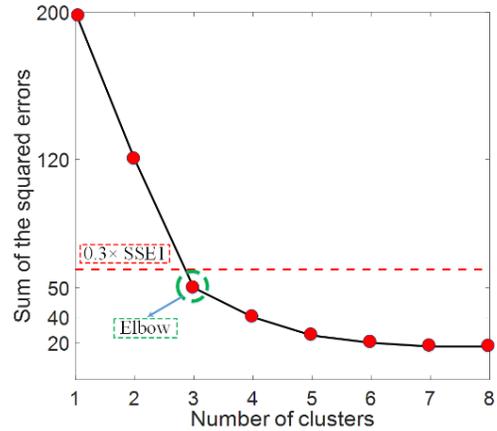

Fig. 3. Elbow point identification.

In the second part, the cross-correlation matrix between nodes with customer loads is visualized in Fig. 6 (b). 0.7 is considered as the limit of the correlation, where a customer load with less than 0.7 correlation with other nodes will be considered in a new group. In the studied network, loads at node 7 and node 38 have low correlation with those in Group 1 and Group 4. Hence, two individual groups (Group 2 and Group 5) are assigned to these two nodes to consider their customer loads behaviours for determining the maximum allowable rooftop PVs. Fig. 5 illustrates the MV distribution network as divided into 5 groups.

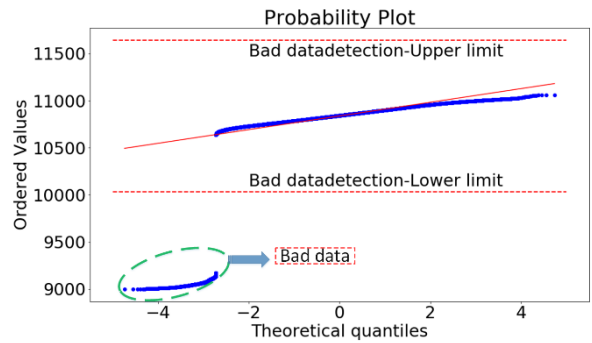

Fig. 4. QQ plot-set of data.

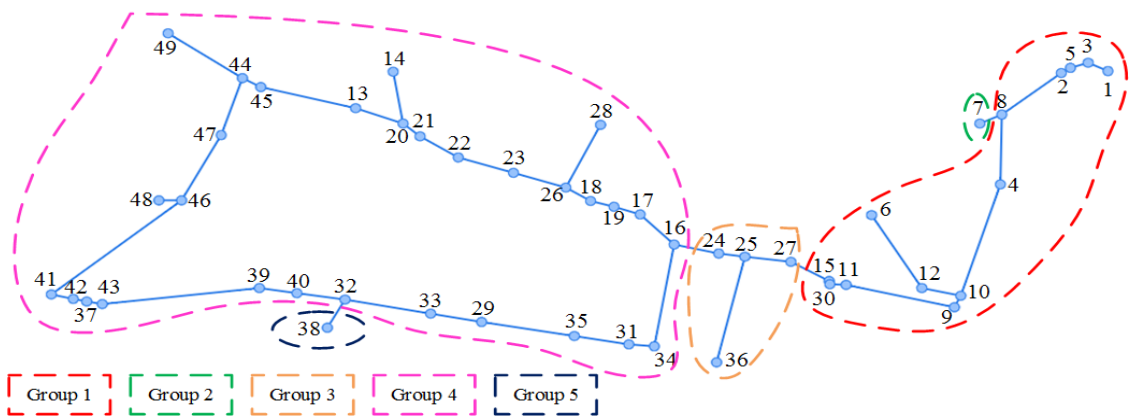

Fig. 5. 49 nodes MV distribution networks.

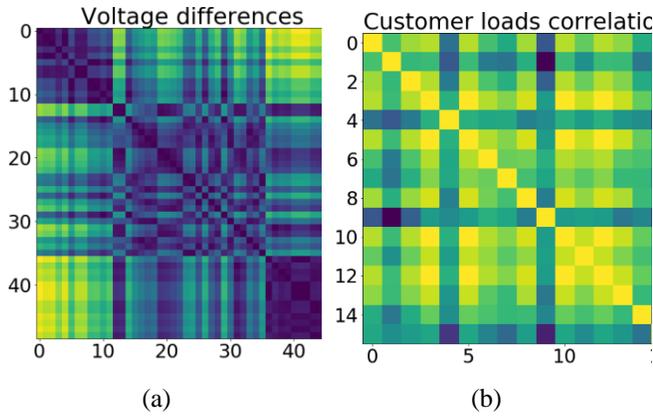

Fig. 6. (a) Nodes voltage differences, (b) Customer loads cross correlation.

*C. Gaussian Distribution Analysis*

Nodal grouping is an important step to put similar neighbouring nodes into the same groups. Based on our investigations, a Gaussian distribution can be fitted on the set of data in each group with a low error. Hence, the voltage of the nodes more than the limit of $\mu + 2\sigma$ provides the time steps that needs to be considered for PV analysis study. Fig. 7 illustrates the result of the fitted Gaussian distribution on the available voltage values in Group 1. The red dashed line represents the $\mu + 2\sigma$.

*D. Two Dimensions K-Means Clustering*

Solar PV's time-varying output is the main reason that studying the critical cases in different time intervals of a day is necessary for finding the maximum allowable PV penetration. In this paper, K-means clustering with Elbow method is employed to partition the available data in each group into the optimal number of the clusters. Fig. 8 represents the result of elbow method for finding the elbow point in Group 1. Three clusters are considered in K-means clustering as shown in Fig. 9. Now, for each cluster, the critical cases can be considered for PV analysis.

For Group 3, the line chart in the Elbow method is smooth in Fig. 10. Hence, the limit $0.3 * SSE1$ is used to find the optimal clusters. As shown in Fig. 11, 4 clusters are considered for Group 3.

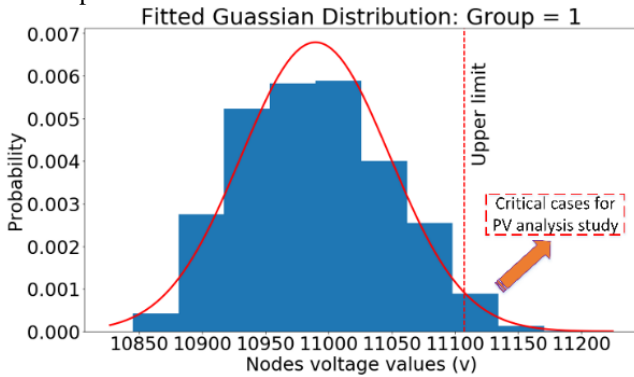

Fig. 7. Fitted Gaussian-set of data in Group 1.

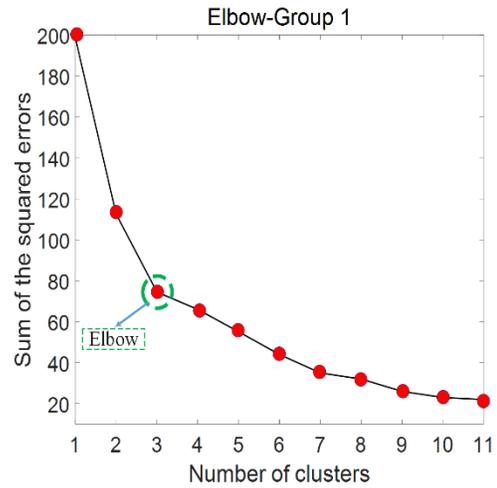

Fig. 8. Elbow point identification-Group 1.

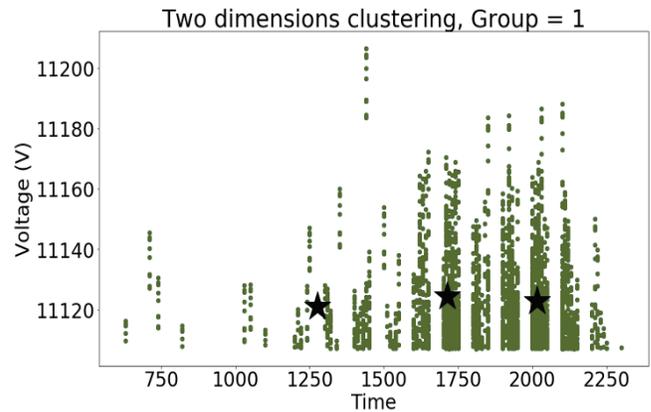

Fig. 9. K-means clustering result-Group 1.

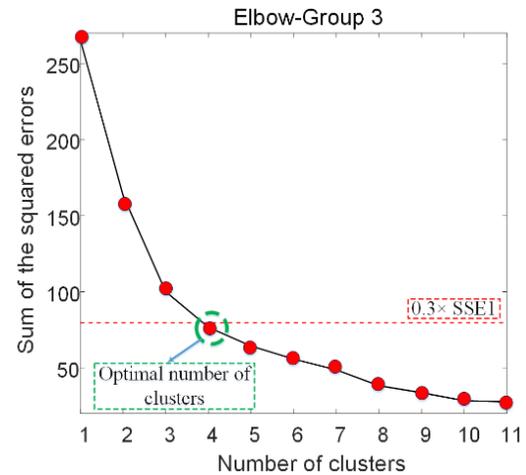

Fig. 10. The optimal number of the required clusters -Group 3.

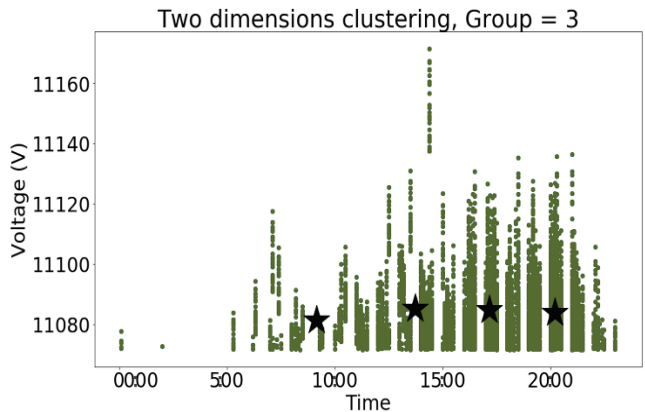

Fig. 11. K-means clustering result-Group 3.

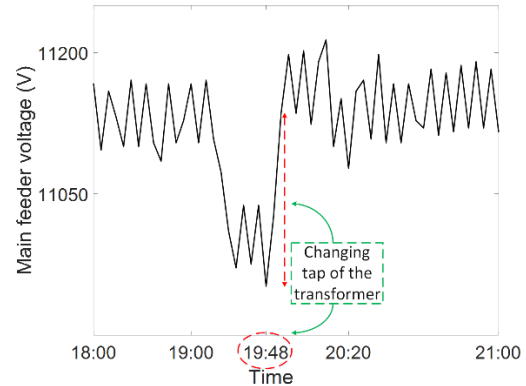

Fig. 12. The main feeder measured voltage data.

The data points in the cluster 3 of Group 1 and cluster 4 of Group 3 are after sunset and do not need to be considered for PV system analysis. It is worth noting that the reason why there are critical cases after sunset is operating with higher transformer's tap in the MV substation. Fig. 12 visualizes the measured voltage data of the main feeder (node 1) of Fig. 1 in January 2017. As shown in this figure, at 19:48, the transformer operated with a higher tap position.

## IV. CONCLUSION

To find the maximum capacity of the lumped rooftop PVs in distribution networks, the ideal scenario would be to study all available estimated and measured data in each season. However, study all available data is impractical for large distribution networks. In contrast to study all available data, only considering the time sample with the highest voltage value cannot be comprehensive enough for rooftop PVs assessment (due to the variability of solar generation). Hence, this paper develops a new data processing approach to optimally identify the critical cases out of the historical data for determining the maximum capacity of lumped rooftop PVs in distribution networks. In the proposed method, bad data are removed and nodal grouping is considered on neighbouring nodes with similar load type. Gaussian distribution is fitted on the nodes voltage values to determine the area with the critical cases, and finally, the optimal number of clusters is considered in two-dimension K-means clustering method to find the critical cases for PV analysis study. The proposed method is employed to find the critical cases out of one complete season data at 10 minutes intervals in an Australian MV distribution network. The proposed method not only detected bad data but also labelled critical cases in each cluster of each group for a further study on the maximum limit of rooftop PVs penetration in the MV distribution network in Brisbane, Australia.

## ACKNOWLEDGEMENT

The authors gratefully acknowledge the support of Australian Renewable Energy Agency (ARENA), Energy Queensland, United Energy, TasNetwork, Australian Power Institute (API), Springfield City Group, the University of Queensland, Queensland University of Technology and Aurecon.